\documentclass[preprint,
 amsmath,amssymb,
 aps,
]{revtex4-1}
\usepackage{graphicx}
\usepackage{dcolumn}
\usepackage{bm}

\usepackage[english]{babel}
\usepackage{upgreek}
\usepackage{setspace}

\setcitestyle{numbers,square}

\usepackage{amsmath,amssymb}
\usepackage{color}
\usepackage{latexsym}
\usepackage[nottoc]{tocbibind}
\usepackage{soul}

\definecolor{dgreen}{rgb}{0,0.5,0}
\definecolor{dred}{rgb}{0.5,0,0}
\definecolor{dblue}{rgb}{0,0,0.5}
\definecolor{dpurple}{rgb}{0.5,0,0.5}

\begin{document}

\preprint{Draft2-RdA-GLT-PDC}

\title{CompuCell3D Model of Cell Migration Reproduces
Chemotaxis}

\author{Pedro C. Dal-Castel$^{1}$, Gilberto L. Thomas$^{1}$, \\ Gabriel C. Perrone$^{1}$, and Rita M.C. de Almeida$^{1,2,3}$
}
\date{February 2021}

\email{pdalcastel@gmail.com}
\email{glt@if.ufrgs.br}
\email{gabriel.perrone@if.ufrgs.br}
\email{rita@if.ufrgs.br}

\affiliation{$^1$Instituto de Física, Universidade Federal do Rio Grande do Sul, Porto Alegre, RS, Brazil}
\affiliation{$^2$Instituto Nacional de Ciência e Tecnologia: Sistemas Complexos, Universidade Federal do Rio Grande do Sul, Porto Alegre, RS, Brazil}

\affiliation{$^3$Programa de Pós-Graduação em Bioinformática, Instituto Metrópole Digital, Universidade Federal do Rio Grande do Norte, Natal, RN, Brazil}

\date{\today}

\begin{abstract}
\vspace{10pt}
\begin{spacing}{1.0}
Chemotaxis combines three processes: directional sensing, polarity reorientation and migration. Directed migration plays an important role in immune response, metastasis, wound healing and development. To describe chemotaxis, we extend a previously published computational model of a 3D single cell, that presents three compartments (lamellipodium, nucleus and cytoplasm), whose migration on a flat surface quantitatively describes experiments. The simulation is built in the framework of CompuCell3D, an environment based on the Cellular Potts Model. In our extension, we treat chemotaxis as a compound process rather than a response to a potential force. We propose robust protocols to measure cell persistence, drift speed, terminal speed, chemotactic efficiency, taxis time, and we analyse cell migration dynamics in the cell reference frame from position and polarization recordings through time. Our metrics can be applied to experimental results and allow quantitative comparison between simulations and experiments. We found that our simulated cells exhibit a trade-off between polarization stability and chemotactic efficiency. Specifically, we found that cells with lower protrusion forces and smaller lamellipodia exhibit an increased ability to undergo chemotaxis. We also noticed no significant change in cell movement due to external chemical gradient when analysing cell displacement in the cell reference frame. Our results demonstrate the importance of measuring cell polarity throughout the entire cell trajectory, and treating velocity quantities carefully when cell movement is diffusive at short time intervals. The simulation we developed is adequate to the development of new measurement protocols, and it helps paving the way to more complex multicellular simulations to model collective migration and their interaction with external fields, which are under development on this date. 
\end{spacing}

\begin{description}
\item[PACS]05.40.-a, 87.17.Aa, 87.17.Jj

\item[Keywords] Single cell directed migration, cell polarization, modified Fürth Equation, CompuCell3D, chemotaxis, chemotactic response, drift speed, chemotactic efficiency
\end{description}
\end{abstract}

\maketitle

\section{Introduction}

\par
Based on their ability to sense environmental cues, migrating cells can follow gradients of temperature, light intensity, electric or chemical fields, of substrate roughness and stiffness \cite{Lauffenburger1996}. This characteristic is determinant for cell survival and large scale organization in multicellular systems. In fact, embryogenesis, inflammatory response, wound healing, and  metastasis depend on the coordination between cells and their environment. Understanding how such phenomena take place, in either computational, experimental or theoretical scopes, requires the consideration of robust measurement protocols and model's faithfulness to reality. We start below by covering important observations and methods of single cell migration without external cues as it is important to support our work on chemotaxis. 

One way to characterize cell movement of isolated mesenchymal cells on isotropic flat surfaces is to calculate Mean Squared Displacement (MSD) as a function of the time interval used to measure cell displacement. MSD curves allow to  identify ballistic and diffusive regimes in different time scales. Fürth equation \cite{Furth20}  has been used to describe single cell migration kinetics for over a century. It describes cell´s MSD with two kinetic regimes: a ballistic regime for short time intervals and a diffusive regime for long time intervals. This equation is, in fact, identical to the solution of the Langevin problem for the velocity of a passive point particle moving in a viscous fluid. In this case, the particle movement at short time intervals is ballistic, therefore instantaneous velocity is a well defined quantity. Opposite to this paradigm, Thomas and collaborators \cite{Thomas2019} analyzed trajectory data from experiments of a cell migrating on flat surfaces and found three different regimes, depending on the time interval. For very short time intervals, the movement is diffusive, for intermediary time intervals the movement is ballistic-like, and for long time scales the movement is diffusive again. Diffusive behavior for short time intervals poses a problem in describing cell movement using instantaneous velocity, since the ratio of cell displacement and time interval diverges if the time interval goes to zero. 

To deal with the short time scale diffusive regime, Thomas and collaborators \cite{Thomas2019} empirically added a term, linear in $\Delta t$, to the Fürth equation, resulting in what they call the modified Fürth equation. This new equation allowed the proposition of natural units for a universal family of curves: cells kinetics only differ by the duration of the short time diffusion interval relative to the onset of the long time interval diffusion. They have successfully fit cells' kinetics for 12 different experimental set-ups from 5 laboratories.  
 
To understand how cells can persistently move while being diffusive in short time scales, we need first to consider cells as active, extensive, irregular and polarized bodies, for which Langevin equations do not apply. Their energy is not harvested from medium thermal activity, being rather based on ATP processing by a complex, internal machinery, where constant polymerization and de-polymerization of actin network takes place, often showing a preferential axis. Cell's center of mass position is the measured quantity to produce  MSD curves and to estimate cell speed. Cell front, or lamellipodium, the main cellular structure responsible for the thrust in cell migration, is very thin as compared to cell body and presents many short lived protrusions in different directions. Each of these protrusions and fluctuations happen simultaneously, at least considering the smallest time intervals available to mesoscopic measurements (of the order of microns). For reviews on cell migration, check Refs. \cite{VicenteManzanares2011,Fortunato2022}. It is reasonable to consider that these fluctuations provide a source of noise to the center of mass position, explaining the short time diffusion observed in single cell migration experiments.
 
The Langevin model can not be directly applied to a system that presents diffusion at short time intervals, since in this case instantaneous velocity is not a measurable quantity. To circumvent this, de Almeida and collaborators \cite{deAlmeida2020} proposed a theoretical, stochastic model of a particle that presents an internally defined polarization. The dynamics of the center of mass position follows a Langevin process in the direction of the polarization and a Wiener process in the polarization' orthogonal direction. The authors have succeeded in obtaining the modified Fürth equation for the MSD (used in Refs. \cite{Thomas2019, Fortuna2020}), with the short time interval diffusion term. Nevertheless, the model failed to obtain the appropriate velocity probability density functions.

In 2020, Fortuna and collaborators \cite{Fortuna2020} developed a simulation model of single cell migration that quantitatively reproduces experimental data. This computational version of a cell is built using Cellular Potts Model (CPM) \cite{Graner1992,Glazier1993,Hogeweg2002} in CompuCell3D \cite{Swat2012}. This model shows cell spontaneous polarization, spontaneous polarization reorientation and migration in the polarization direction. The resulting MSD curves are well fit by the modified Fürth equation, and Mean Velocity Correlation Functions (mVACF) reproduce the experimental results. The fitting procedure, using the modified Fürth equation, allowed a translation between laboratory and simulation units. In fact, the model serves as a proxy to the experimental cell behavior and provides a powerful tool to investigate cell migration kinetics, to be verified later in experiments. Thomas et. al. \cite{Thomas2022} investigated the polarization definition in this model, verified that the dynamics of cell migration is different in the polarization direction and orthogonal direction, and proposed a polarization measurement procedure that also applies to experiments. 

Here we present a modification of Fortuna and collaborator's model to simulate chemotactic response, and we provide an appropriate MSD equation to characterize cell kinetics in this condition. We aim at a chemotactic response mechanism focused on migration reorientation, agreeing with experimental evidence that external chemical gradients act primarily over polarization orientation rather than cell speed \cite{Korohoda2002, Neilson2011May}. Then we further investigate cell polarization and the diffusive regime for short time scales to propose new metrics. Finally, we apply both new and standard metrics to characterize the response and compare it to the non chemotactic, isotropic case. 

The paper is organized as follows. In Section \ref{MATMET} we present the simulation model, discuss how we implement the response to external gradients, and propose adequate measures to investigate both the cellular response and the short time interval behavior with and without external fields. In Section \ref{RESULTS} we present the results and in Section 4 we discuss and concludes with new ideas for experiments.


\section{Methods}
\label{MATMET}

\subsection{Simulation Model}
\label{Model}

\subsubsection{Original Model}

Fortuna and collaborators proposed a simulation model that describes single cell migration on flat substrates \cite{Fortuna2020}. The model is based on CPM \cite{Graner1992,Glazier1993,Hogeweg2002} and is  built in the CompuCell3D (CC3D) environment \cite{Swat2012}. The simulation consists in a three dimensional grid containing the migrating cell immersed in a Medium and laying over a 2D flat Substrate. The simulated cell contains three objects: Lamel, Nuc, and Cyto (simulated versions of the real lamellipodium, nucleus and cytoplasm, respectively). Lamel is a structure that can protrude towards the Medium due to an internal F-actin field. In the lattice, an object is a set of voxels (3D pixels) with same labels. The model assigns two labels to each lattice site: $\sigma$ indicates the cell and $\tau$ discriminates compartments. This way, different parts of the same cell may show different behaviors. A global energy $E$ is attributed to the lattice configuration. The algorithm randomly chooses a pair of neighboring sites and the labels of the first site of the pair is tentatively copied over the second site. When this change decreases the system energy, the copy is accepted. When this change increases the energy by $\Delta E$, the copy is accepted with probability 

\begin{equation}
P_{Boltzmann}=e^{-\frac{\Delta E}{T_B}} \;\;\; ,
\label{eq:ProbBoltz}
\end{equation}
where  $T_B$ is a Boltzmann-like temperature parameter, associated to membrane fluctuations \cite{Swat2012}. After this process, a new pair of sites is chosen and the same routine is repeated. A Monte Carlo Step (MCS) is defined as $N_{MCS}$ repetitions of the above process, where $N_{MCS}$ equals the total number of voxels in the lattice.

The model proposed by Fortuna and collaborators \cite{Fortuna2020} considers that the energy of a given configuration is the sum  of contact and volume terms, such that

\begin{equation}
\label{Eq:Etotal}
    E_{total} = E_{contact}+E_{volume} \;\;\; ,
\end{equation}
where
\begin{equation}
\label{Eq:Econtact}
    E_{contact}=\sum_{\vec{i}}\sum_{\langle \vec{j}\rangle _{\vec{i}}}J(\tau(\vec{i}),\tau({\vec j})) \;\;\; ,
\end{equation}
and
\begin{equation}
\label{Eq:Evolume}
    E_{volume}=\sum_{\sigma}\lambda_{\sigma}(V_{current}^{\sigma}-V_{target}^{\sigma})^2 \;\;\; .
\end{equation}
Here, $J(\tau(\vec{i}),\tau(\vec{j}))$ is the energy per edge between neighboring sites $\vec{i}$ and $\vec{j}$, each with different labels $\tau$. The expression $\langle \vec{j}\rangle _{\vec{i}}$ represents the neighboring sites of $\vec{i}$. For neighboring sites of same type, contact energy is set to zero. $\lambda_{\sigma}$ is the inverse of $\sigma^{th}$ cell's compressibility, $V_{current}^{\sigma}$ and $V_{target}^{\sigma}$ are, respectively, its current and target volumes. These energy terms jointly play an important role in cell organization, shape and size.

\begin{figure}
    \centering
    \includegraphics[scale=0.7]{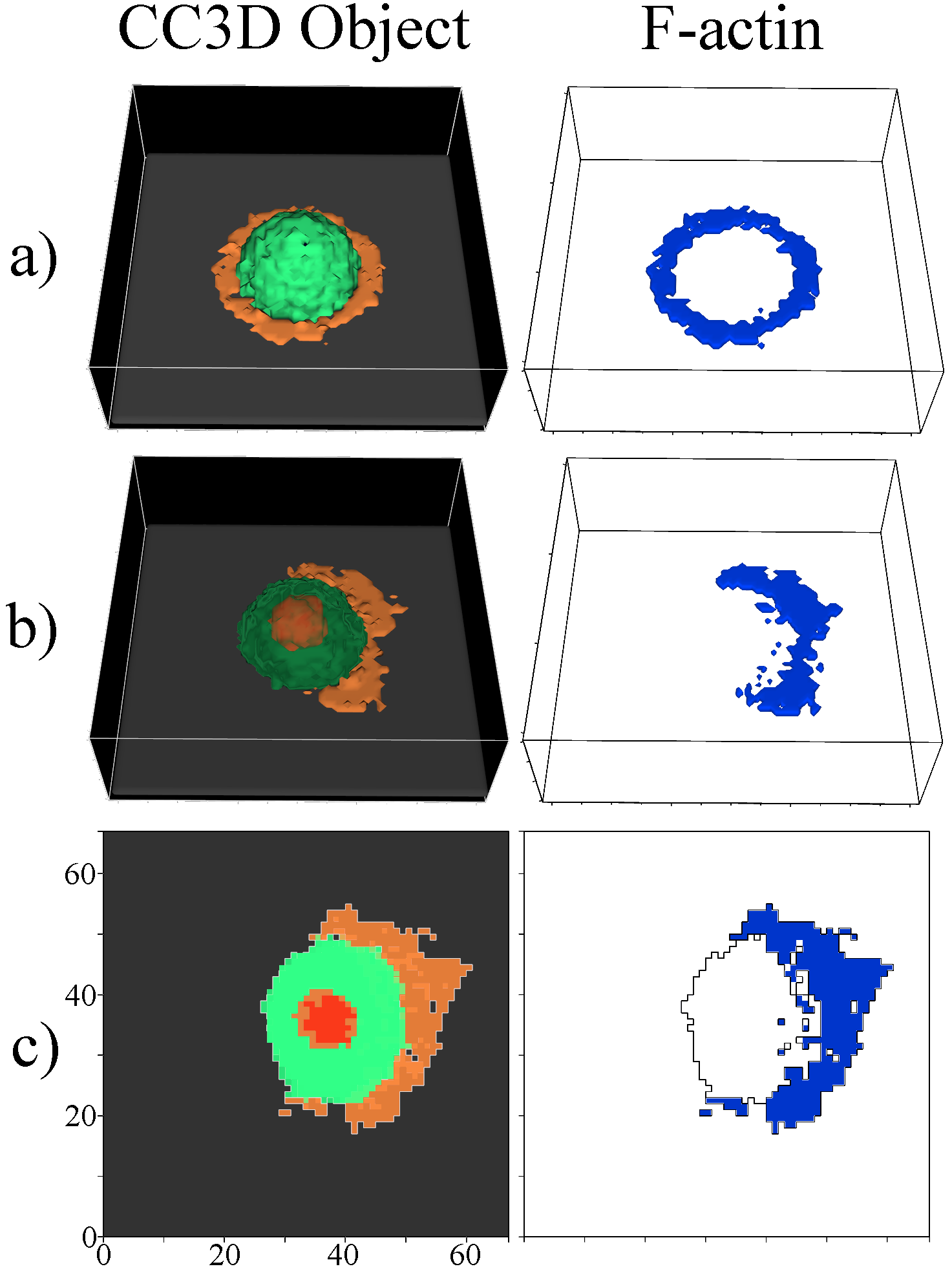}
    \caption{Visualization of the CC3D objects composing the simulated cell and the internal field F-actin. \textbf{a)} The cell initialization is a symmetrical semi-sphere over the flat substrate. \textbf{b)} Cell in a polarized state. Image was edited to show the Nuc inside. \textbf{c)} Cell in a polarized state (view of three 2D layers superposed in the xy plane).  In all cases, the internal field F-actin has value 1 (blue) in the Lammelipodium voxels and 0 (white) otherwise. Note that, in the polarized state, the Nuc lags behind inside the Cyto, enabling the definition of a polarization vector in Eq. \ref{eq:POL}, which we will use later to define a cell reference frame.}
    \label{fig:CELL}
\end{figure}

Lamel protrusions are the motors of cell migration. This action can not be described in CPM by a potential energy. Fortuna and collaborators achieved a protrusion behavior via an energy variation term $\Delta E_{protrusive}$, calculated only for Lamel voxels in contact with the substrate. When a copy of a Lamel site $\vec{j}_{Lamel}$ over a medium site at $\vec{i}_{Medium}$ is attempted, an additional energy term is calculated and added to the energy change, simulating the work done by the cell's non-conservative internal processes. This term is  
\begin{equation}
\label{Eq:Eprotrusive}
  \begin{array}{l}
    \Delta E_{protrusive} = -\lambda _{F-actin}[F(\vec{i}_{Medium})-F(\vec{j}_{Lamel})] \; \; \; ,
  \end{array}
\end{equation}
where  $F(\vec{i}_{Medium})-F(\vec{j}_{Lamel})$ represents an actin field gradient between the neighboring voxels of Medium and Lamel at sites $\vec{i}$ and $\vec{j}$, respectively. $F$ is a discrete field set equal to one in Lamel voxels and to zero otherwise. This dynamics favors Lamel growth over Medium rather than over Cyto. Then, a backpropagation of volume interaction takes place: first, the Lamel volume increases due to protrusions, then Cyto grows towards Lamel in an attempt to balance volume energies, which, in turn, favors Medium growth over Cyto in the rear. Finally, the Nuc lags behind, but its high contact energy with Medium pushes it forward. As a consequence of this process, the entire cell moves. Fortuna and collaborators finely tuned the parameters to prevent separation of Lamel and Cyto and other artifacts. This dynamics makes the cell a self polarizing structure with a preferential migration axis that changes direction over time. Furthermore, it mimics a Local Excitation - Global Inhibition (LEGI) dynamics \cite{Turing1952, Swaney2010, Weiner2002, Chung2002, Manahan2004}: the more Lamel at one region (the more Lamel/Medium interface), the larger the probability of increasing Lamel at that region. The consequent increase in Lamel, on the other hand, decreases the probability of growing Lamel everywhere else. Fig. \ref{fig:CELL} shows the cell structure and internal F-actin field in 3 different perspectives. For more details, see Refs. \cite{Fortuna2020}.

Cell kinetics resulting from the  model described in Ref. \cite{Fortuna2020} agrees with experiments for cells in the absence of external fields. The model, however,  does not have any mechanism to promote cell response to environmental cues. In the next topic, we show an adaptation of this simulation to model chemotactic response.

\subsubsection{Modified Model: Chemotaxis}

Eukaryotic cell migration requires cell cytoskeleton organization. In several eukaryotic cell species, $PI_3K$, $PIP_3$, Rac and Rho GTPases implement a Local Excitation Global Inhibition (LEGI) dynamics \cite{Turing1952,Swaney2010, Weiner2002, Chung2002, Manahan2004} that regulates lamellipodium polimerization and depolimerization. This way, a large and localized lamellipodium promotes its own localized growth, while inhibiting growth in other locations. For more details, see Ref. \cite{VicenteManzanares2011}.

In chemotaxis, some chemical concentration around the cell presents a gradient that is spatially sensed by the cell. This signal is biochemically transduced and perturbs cell's internal machinery, favoring a given orientation for lamellipodium growth, directing migration. Here we adapt the model to simulate this three-step dynamics (sensing, reorientation and migration \cite{Roussos2011,Jin2008}). 

\begin{figure}
    \centering
    \includegraphics[scale=0.5]{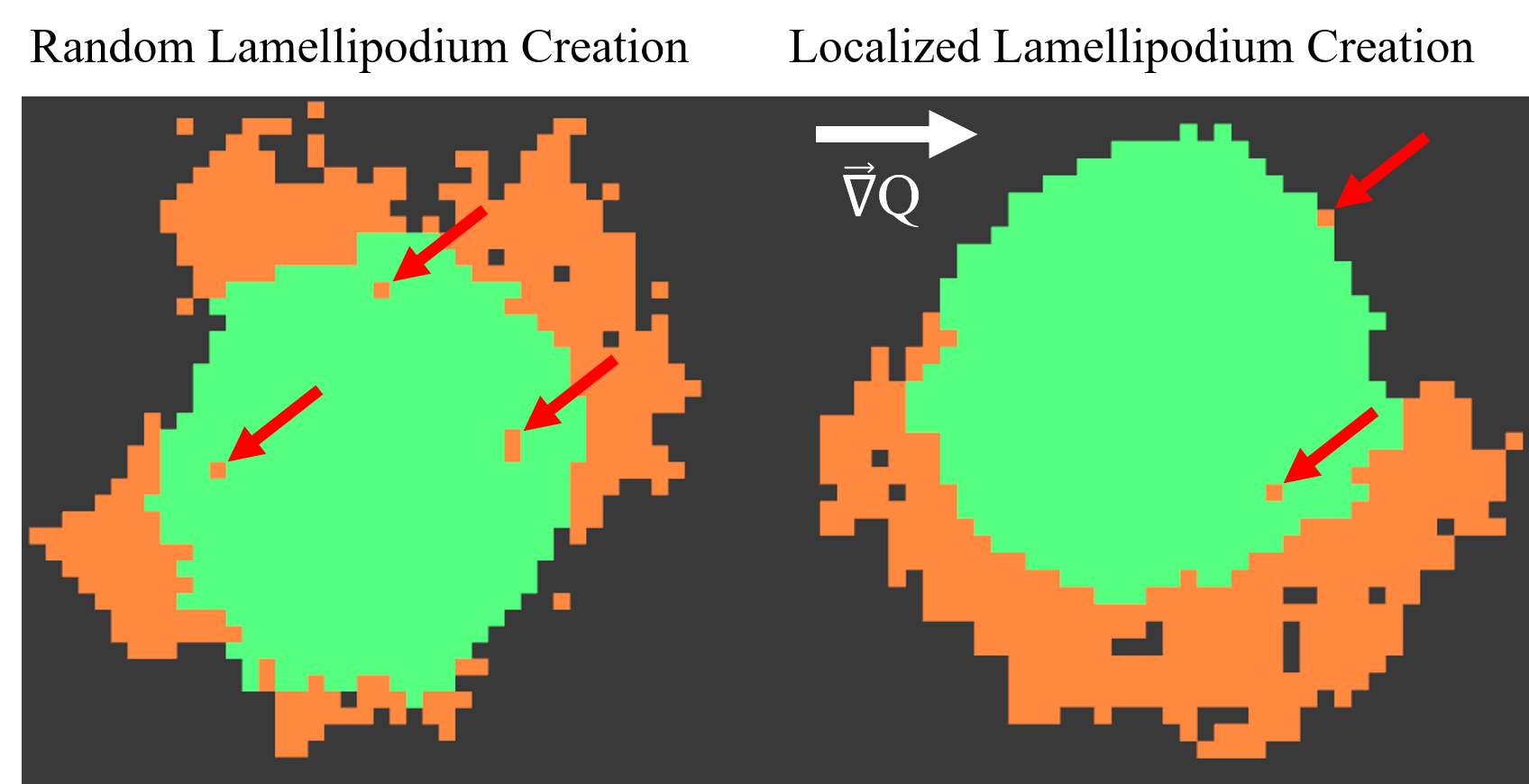}
    \caption{In the left, the cell does not sense any environmental anisotropy, so Lamel voxels are created randomly as tells Eq. \ref{Eq:Pcytolamel} for $Q(\vec{i})-\langle Q\rangle_{cell}=0$. In the right, the cell sensing an external field gradient will create Lamel asymmetrically, following the same equation \ref{Eq:Pcytolamel}, directing its migration in the long run. The red arrows point to the newly created Lamel (orange) voxels, the white arrow indicates the chemical gradient direction. The images are 2D slices at z=1, where the voxel conversion takes place.}
    \label{fig:Cell_Response}
\end{figure}

We define a linear constant external chemical field $Q(\vec{i})$, whose concentration is sensed in the cell base. By comparing local concentrations to the average concentration sensed by the cell, the cell creates new Lamel voxels in higher concentration sites, as we outline in Fig. \ref{fig:Cell_Response}. This creation of Lamel increases stability of polarization in the direction of the gradient, which then directs migration. With this dynamics, we aim at a chemotaxis response by reorientation (sometimes referred as compass model \cite{Bourne2002, Neilson2011May}) rather than an explicit force acting over the cell. The implementation of the following dynamics in CC3D can be accessed and downloaded from GitHub \cite{Pedro2023}.


To control the production of Lamel voxels in response to chemotaxis, we only allow the creation of new Lamel voxels when Lamel's current volume is below the Lamel average volume taken over the last 100 MCS. This memory mechanism can be expressed by a switch function 
\begin{equation}
    Switch = \frac{\langle \phi _f \rangle _{n-100}^n - \phi _f ^n}{|\langle \phi _f \rangle _{n-100}^n - \phi _f ^n|} \;\;\; ,
    \label{eq:switch}
\end{equation}
where $ \phi _f^n $ is Lamel's volume fraction (relative to cell's volume) at time step $n$, and $\langle \phi _f \rangle _{n-100}^n$ is the average of $ \phi _f^n $ over the previous 100 steps. With this memory mechanism active, cell's chemotactic response persists in time. In real cells, actin polymerization is backed by time persistent auxiliary mechanism such as cytoskeleton polarization and non uniform distribution of myosin, integrins and several migration signals within the cell \cite{CallanJones2016}. These intracellular processes allow persistent cell migration. Our choice of 100 MCS is an intermediate value between the regime of pure Potts fluctuations (10 MCS or lower) and the regime of cell migration persistence (1000 MCS or higher). 

Merging the sensing and the memory mechanisms in a single equation, the probability $P_{convert}$ of a Cyto's voxel conversion to Lamel is given by

\begin{equation}
\label{Eq:Pcytolamel}
    P_{convert}(\vec{i})=\rho \; \left [ \tanh \left( \mu \frac{Q(\vec{i})-\langle Q\rangle_{cell}}{\sigma _{Q,cell}}\right)+1\right ]\times Switch\; . 
\end{equation}


The first factor on the r.h.s. accounts for the directionality in the Lamel creation at Cyto's base. The external field $Q(\vec{i})$ is evaluated at every voxel $\vec{r}$ at the cell's base and $\langle Q\rangle_{cell}$ is its spatial average. The difference between the value of $Q(\vec{i})$ and the field's average determines the likelihood that a Cyto's voxel will switch to Lamel at position $\vec{i}$. The $\mu$ parameter is the hyperbolic tangent steepness at $Q(\vec{i})-\langle Q\rangle_{cell}=0$, it regulates the asymmetry in Lamel voxel creation. The argument is normalized by the standard deviation of the field $\sigma _{Q,cell}$ so that $\mu$ stands for both field gradient intensity and cell's susceptibility. At last, $\rho$ normalizes the probability, so it assumes value $\rho=1/2$. The second factor is the switch we defined in Eq. \ref{eq:switch}. Both factors together regulate directionality of Lamel creation and persistent response through time. 
After sensing and reorienting, the F-actin promoted protrusions in Lamel accomplishes the third step of the chemotactic response, i.e., migration. This model achieves chemotactic response as show the trajectories in Fig. \ref{fig:msdfig} in Section \ref{TRAJ}. 

\subsubsection{Simulation Execution Time}
The simulation execution time, in a Potts simulation, typically scales with the number of lattice sites $N_{voxels}=L_x \times L_y \times L_z$, but can change depending on plugins and routines used. In our simulations, $N_{voxels}$ depends on the cell size, because we adjust the lattice to the cell. One simulation with a single cell of radius 10 and 100000 MCS, on an Intel i7-3770K processor takes approximately 4 hours.


\subsection{Quantitative Characterization of Cell Movement}
A common experimental set-up for studying  cell migration consists in cultivating cells over adherent flat substrates in a Petri dish. For low cell density, single cell migration takes place and chemical gradients can be imposed with external control \cite{Korohoda2002,Metzner2021, Neilson2011May,Tweedy2013Sep}. Cells are monitored by microscope, their positions and shapes are tracked by time-lapse imaging. Cell's internal chemical species such as ERK, F-actin, and membrane receptors can also be monitored using fluorescence techniques \cite{Devreotes2003,Cao2004,RocaCusachs2013}. After cell trajectories are obtained, it is possible to characterize cell movement.

\subsubsection{Mean Square Displacement - MSD}
\par
A plot of the MSD versus time interval $\Delta t$ shows different movement regimes indicated by its slope in a log-log scale. Ballistic regimes have slope equal to $2$ while diffusive regimes have slope equal to $1$. The mathematical definition of the MSD is

\begin{equation}
	\langle  |\Delta\vec{ r}|^2 \rangle = \frac{1}{M} \sum_{M}\; \frac{1}{N-\Delta t}\; \sum_{i=1}^{N-\Delta t}|\vec{r}(t_i+\Delta t)-\vec{r}(t_i)|^2\;\;\; ,
	\label{eq:MSD}
\end{equation}
where $M$ is the total number of different $N$ steps trajectories, $\vec{r}(t)$ is the cell position at time $t_i$ (subscript $i$ indicates the discrete counting of time: $t_1$, $t_2$...). $\langle  |\Delta\vec{ r}|^2 \rangle$ is a function of $\Delta t$ - the time interval between two cell positions - and it is averaged over all points of a trajectory and all acquired trajectories. MSD curves can be fit using models such as the Fürth equation \cite{Furth20}, from which it is possible to calculate persistence and diffusivity. In the Supplementary Materials \cite{Pedro2023supp}, we show a step-by-step fitting process of a MSD curve with 4 regimes of movement.

\subsubsection{Mean Velocity and Optimal Velocity}
\par
Normally, we would consider instantaneous velocity as another possible characterization measurement, but cells can present a diffusive behavior for short time scales \cite{Thomas2019}. Instead, we measure mean velocity using
\begin{equation}
 \vec{V}(t,\delta) = \frac{\vec{r}(t_i+\delta)-\vec{r}(t_i)}{\delta} \;\;\; ,
	\label{eq:mVEL}
\end{equation}
where $\delta$ is the time interval between two cell positions. 

It is possible to define an optimal mean velocity $\vec{V}_{opt}=\vec{V}(t_i,\delta _{opt})$, where $\delta _{opt}$ is the time interval at which MSD slope is the steepest. If the cell is ballistic for short time scales, $\vec{V}_{opt}$ will be taken as the limit of $\vec{V}$ for $\delta \rightarrow 0$, as normally used.

\subsubsection{Mean Velocity Autocorrelation Function (mVACF) and Displacement Autocorrelation Function}
\par 
An important measurement that considers cell mean velocity is Mean Velocity Autocorrelation Function - mVACF. mVACF can be used to calculate a characteristic memory time and to determine underlying dynamics \cite{Vinales2020}. It is defined as 

\begin{equation}
	\text{mVACF} = \frac{1}{M} \sum_{M}\; \frac{1}{N-\Delta t - \delta}\; \sum_{i=1}^{N-\Delta t-\delta}\vec{V}(t_i,\delta)\cdot \vec{V}(t_i+\Delta t,\delta) \;\;\; ,
	\label{eq:mVACF}
\end{equation}
where $\vec{V}(t_i,\delta)$ follows the definition of mean velocity in Eq. \ref{eq:mVEL}. mVACF remains as a function of $\Delta t$ - the time interval between two velocity measurements -  and $\delta$ - the time interval used to measure mean velocity in Eq. \ref{eq:mVEL}. Note that $\Delta t > \delta$ so that two successive mean velocity measurements do not overlap in time, which would introduce an artificial correlation \cite{Thomas2019}. The first average is taken over all $N$ points of the trajectory, except when $t_i>N-\Delta t - \delta$, for which the next mean velocity measurement would fall outside the range of trajectory points. Another average is taken over all trajectories at hand, analogous to MSD.

Since cells are often polarized structures, another autocorrelation to consider is of cell displacement in the directions parallel and perpendicular to polarization. This metric extract further information about the sistem's underlying dynamics. Using the symbols $\langle \rangle$ to denote ensemble averages, we define the displacement autocorrelation function as
\begin{equation}
    C_{rr} = \langle (\Delta \vec r _{t}\cdot \vec\Pi_t)(\Delta \vec r _{t+\Delta t}\cdot \vec\Pi_{t+\Delta t}) \rangle _{\xi}
\end{equation}
for the direction parallel to the polarization $\vec\Pi$, defined in Eq. \ref{eq:POL} (see following subsection). For perpendicular direction, $C_{rr}$ is analogous, but considers $\vec\Pi$ rotated $90^{\,\small{\mbox{o}}}$ in the $xy$ plane.

\subsubsection{Polarization and Polarization Direction Distribution}
\par
Polarization is the preferential direction of internal cellular fibers that participate in the transport of the necessary molecules to build the actin network at cell front and to retract it at cell rear. This direction correlates to cell's drift speed \cite{CallanJones2016}. In our simulations, due to cell movement, Nuc tends to lag behind of cell's geometrical center. As in Ref. \cite{Thomas2022}, we take advantage of Nuc localization to define polarization as

\begin{equation}
\vec{\Pi}=\vec{r} _{\text{CN}}-\vec{r} _{\text{N}} \;\;\; ,
	\label{eq:POL}
\end{equation}
where $\vec{r} _{\text{N}}$ is Nuc's center of mass position and $\vec{r} _{\text{CN}}$ is the center of mass position of Nuc and Cyto combined. This definition leads to a vector correlated with cell displacement \cite{Thomas2022}. Experimentally, this definition  requires the nucleus and the cell border to be visible throughout the experiment. Chemical species distribution inside the cell are also a possible asset for determining cell polarity.

If the cell is responding to an external chemical gradient, the histogram of polarization angle with the $x$-axis of the laboratory reference frame, $\theta$, is a possible measurement of cell's orientated migration. A directional cell movement will lead to a spike in the distribution of $\theta$ at the direction of the gradient,  while a cell movement in the absence of chemical gradients leads to a flat distribution of $\theta$. 

\subsubsection{Drift Speed}
\par
We define drift speed as the average over time and trajectories of the mean velocity projected on the polarization vector for $\delta \rightarrow 0$:
\begin{equation}
	V_d = \frac{1}{M} \sum_{M}\;\; \lim_{\delta\to 0} \frac{1}{N-\delta}\; \sum_{i=1}^{N-\delta} \frac{\vec{V}(t_i,\delta)\cdot\vec{\Pi} _i}{|\vec{\Pi} _i|} \;\;\; .
	\label{eq:VD}
\end{equation}
$\vec{\Pi} _i$ is the polarization at time $t_i$ defined in Eq. \ref{eq:POL}, $\vec{V}(t_i,\delta)$ is the mean velocity defined in Eq. \ref{eq:mVEL}.  The resulting scalar $V_d$ measures how fast the cell moves in the polarization direction in average, which may converge for small time intervals $\delta \rightarrow 0$ even if the movement is diffusive for short time intervals. That could happen when the diffusive contribution for displacement has zero average. Suppose the cell displaces $\Delta \vec r_d$ in the direction of polarization due to the drift plus a random diffusion $\Delta \vec\xi$ during a time interval equals to $\delta$. Then,
\begin{equation}
    \left \langle \vec V (t_i, \delta) \cdot \frac{\vec{\Pi} _i}{|\vec{\Pi} _i|} \right \rangle = \left\langle \left (\frac{\Delta \vec r_d}{\delta} + \frac{{\Delta \vec\xi}}{\delta} \right ) \cdot \frac{\vec{\Pi} _i}{|\vec{\Pi} _i|} \right\rangle = \left\langle \frac{\Delta \vec r_d}{\delta}  \cdot \frac{\vec{\Pi} _i}{|\vec{\Pi} _i|} \right\rangle\;\;\; ,
\end{equation}
since the noise term average is zero.
The average operation $\langle \; \rangle$ allows the convergence of $\left \langle \vec V (t_i, \delta) \cdot \frac{\vec{\Pi} _i}{|\vec{\Pi} _i|} \right \rangle$ even though $\frac{{\Delta \vec\xi}}{\delta}$ diverges for $\delta \rightarrow 0$. Consequently $V_d$ is well defined. 

If the cell responds to an external chemical gradient but no significant change in drift speed is observed, it must be the case that cell's polarization is aligned to the gradient. That is why analyzing both polarization angle distribution (previous item) and drift speed is important to fully understand how cells respond to stimuli.

\subsubsection{ Terminal speed and the Chemotactic Efficiency}
\label{chemeffsec}

When cells respond to external chemical gradients, a non zero terminal speed appears, causing a MSD curve with slope equals to $2$ for large $\Delta t$. F. Peruani and L. G. Morelli \cite{Peruani2007} analytically demonstrated that an orientation directionality only (without changes in speed behavior) causes an extra ballistic term for long time intervals in the MSD. Therefore, we approximate the MSD in this regime to
\begin{equation}
     \langle  |\Delta\vec{ r}|^2 \rangle \sim B \Delta t^2 \;\;\; ,
\end{equation}
where $V_T=\sqrt{B}$ is the terminal speed of the cell (more details in the results section). We define the ratio between terminal speed and drift speed (defined in Eq. \ref{eq:VD})
\begin{equation}
    \varepsilon = \frac{V_T}{V_d} \;\;\; ,
    \label{Eq:ChemEFF}
\end{equation}
as a measurement of chemotactic efficiency. The idea behind this metric is that a cell perfectly aligned to the chemical gradient will convert all its drift speed into terminal speed, whereas a non perfectly aligned cell will certainly have a terminal speed smaller than its drift speed. Provided that the cell has a net velocity (drift speed) in the direction of polarization, this metric can be applied to any taxis mechanism, since every directed migration will appear as a long term ballistic MSD curve.

\subsubsection{ Cell Displacement Distributions and  Cell Reference Frame }

The investigation of cell movement in respect to the laboratory reference frame and to the cell reference frame requires a detailed set of measurements. Here we present the distribution of the cell displacement for a given time interval $\Delta t$ as our final way to quantify and visualize the cell behavior in the 2D plane. To perform this metric, we 
\begin{enumerate}
    \item take the cell's trajectory with the polarization vector in each time step; 
    \item choose a set of time intervals $\Delta t$ to analyze; 
    \item calculate the tuples $(\Delta \Vec{r}_x, \Delta \Vec{r}_y)$ and $(\Delta \Vec{r}_\perp, \Delta \Vec{r}_\parallel)$ for all time points and trajectories and every chosen $\Delta t$, where the last tuple is the parallel and perpendicular component of the displacement relative to the polarization vector measured in the beginning of each step $\Delta t$;
    \item perform the 2D frequency counts of the tuples of $\Delta \Vec{r}$ for the two coordinate systems separately; and
    \item plot resulting 2D histograms.
\end{enumerate}

When  displacement is projected onto polarization direction, a new reference frame emerges, where the $x$-coordinate is the displacement component in the direction of polarization and the $y$-coordinate is the displacement component in the perpendicular direction of polarization. This strange non-inertial reference frame allows us to further comprehend cell migration dynamics, as it is inextricably linked to polarization.


\section{Results}
\label{RESULTS}

\subsubsection{Varying cell radius, Lamel volume and protrusion strength modifies cell migration.}

To characterize cell movement, we measured the mean velocity components relative to polarization and we compare the average of mean velocity with the average of absolute mean velocity for all available time intervals $\Delta t$.

We varied three parameters, as shown in Fig. \ref{fig:velfig}. First, we observed that drift speed increases with $\lambda_{F-actin}$ - the protrusion coefficient. This happens because $\lambda_{F-actin}$ regulates the likelihood of a Lamel voxel to be copied over a Medium voxel. If $\lambda_{F-actin}$ is too high ($\lambda_{F-actin}> 175$), unwanted effects start to take place, such as Lamel breaking from Cyto. Second, we observe that drift speed decreases for $\phi_f$ too large or too small. For too large $\phi_f$, Lamel organizes as a ring around the cell, disfavoring polarization. In the opposite case, if $\phi_f$ is too small, there will not be enough Lamel to protrude, reducing drift speed. At last, cell radius changes the relative effect of the membrane fluctuations (regulated by $T_B$, see Eq. \ref{eq:ProbBoltz}) and Lamel's size. For larger cell radius of $R=15$ and $R=20$, the area covered by Lamel is comparatively larger than $R=10$ because $\phi_f$ grows with $R^3$ while Lamel is flat, and the average membrane fluctuation will have a reduced comparative effect over the larger cells because it only acts in the interface between two CPM objects. These factors imply that to correctly scale the cell by radius, other parameters must scale together. Next results will be restricted to $R=10$.

\subsubsection{Short time diffusive behavior is isotropic}
\label{velocity1}

\begin{figure}{htbp}
    \centering
    \includegraphics[scale=0.75]{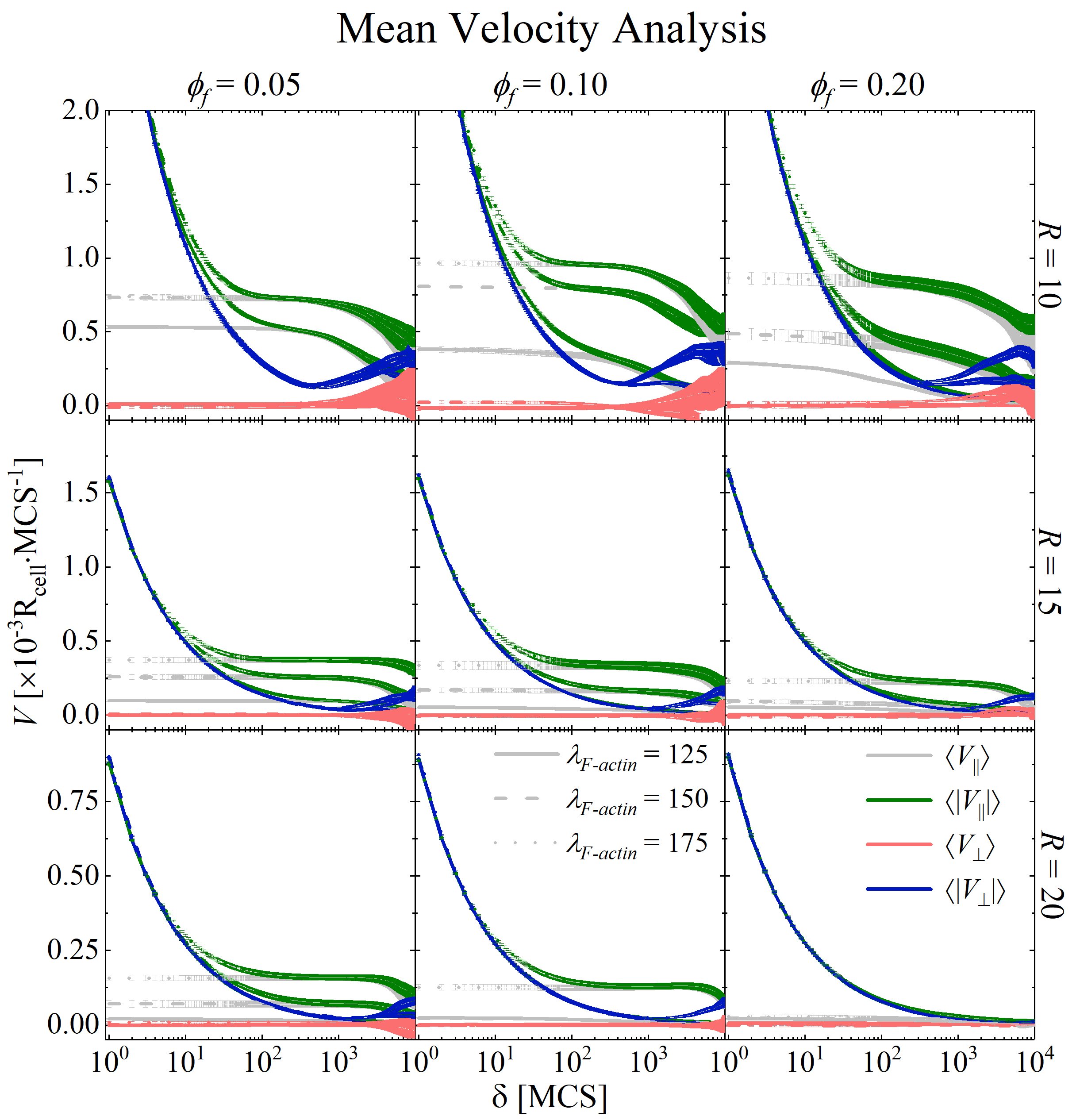}
    \caption{{\setstretch{1.0} Four different measurements as functions of $\delta$, the time interval used to calculate the mean velocity: \textbf{1)} $\langle V_\parallel \rangle$ is the average of the parallel component of the mean velocity with respect to the polarization vector (gray); \textbf{2)} $\langle |V_\parallel |\rangle$ is the same, but the absolute value is taken before averaging (green); \textbf{3)} $\langle V_\perp\rangle$ is the average of the perpendicular component of the mean velocity with respect to the polarization vector (red); \textbf{4)} $\langle |V_\perp|\rangle$ is the same, but the absolute value of velocity is taken (blue). We measured mean velocity for three different values of relative Lamel volume $\phi _f$ (graph columns), Lamel protrusion coefficient $\lambda _{F-actin}$ (different line styles), and different cell radius (graph rows). Divergence in $\langle |V_\parallel |\rangle$ and $\langle |V_\perp|\rangle$ as $\delta \rightarrow 0$ demonstrates the isotropic diffusive noise, unlike $\langle V_\parallel \rangle$ and $\langle V_\perp\rangle$, which converge to $V_d$ and zero, respectively. Following Eq. \ref{eq:VD}, we extract drift speed $V_d$ from the graphs by taking the $\langle V_\parallel \rangle$ value for the smallest $\delta$, since it converges.}}
    \label{fig:velfig}
\end{figure}

The simulated cell has a polarization direction defined by Eq. \ref{eq:POL}, with distinct dynamics for parallel and perpendicular directions in respect to it. By projecting cell's mean velocity in both perpendicular $(V_\perp)$ and parallel $(V_\parallel)$ directions to the cell polarization at the beginning of the time interval, we find that a diffusive noise is present in both directions, shown by the divergence in the modules of both velocity components. Fig. \ref{fig:velfig} shows $\langle |V_\perp |\rangle$ and $\langle |V_\parallel |\rangle$ as functions of the time interval $\delta$ used to calculate mean velocities for different simulation parameters. We observe that both quantities diverge as $\Delta t \rightarrow 0$, implying that instantaneous velocity can not be defined in any direction.


\subsubsection{The modulus of finite drift speed is not significantly affected by the external field }

To investigate the effect of an external chemical field \textbf{$Q(\vec{r})$} (constant in time, linear in $x$-direction), we set a saturation value $\mu=10^{6}$ for Eq. \ref{Eq:Pcytolamel}. We measure mean velocity, same as in Section \ref{velocity1}, the distribution of polarization angle $\theta$, and we compare the sensitive cell $\mu=10^{6}$ subjected to the gradient to the isotropic case (no external chemical gradients).

\begin{figure}{htbp}
    \centering
    \includegraphics[scale=0.7]{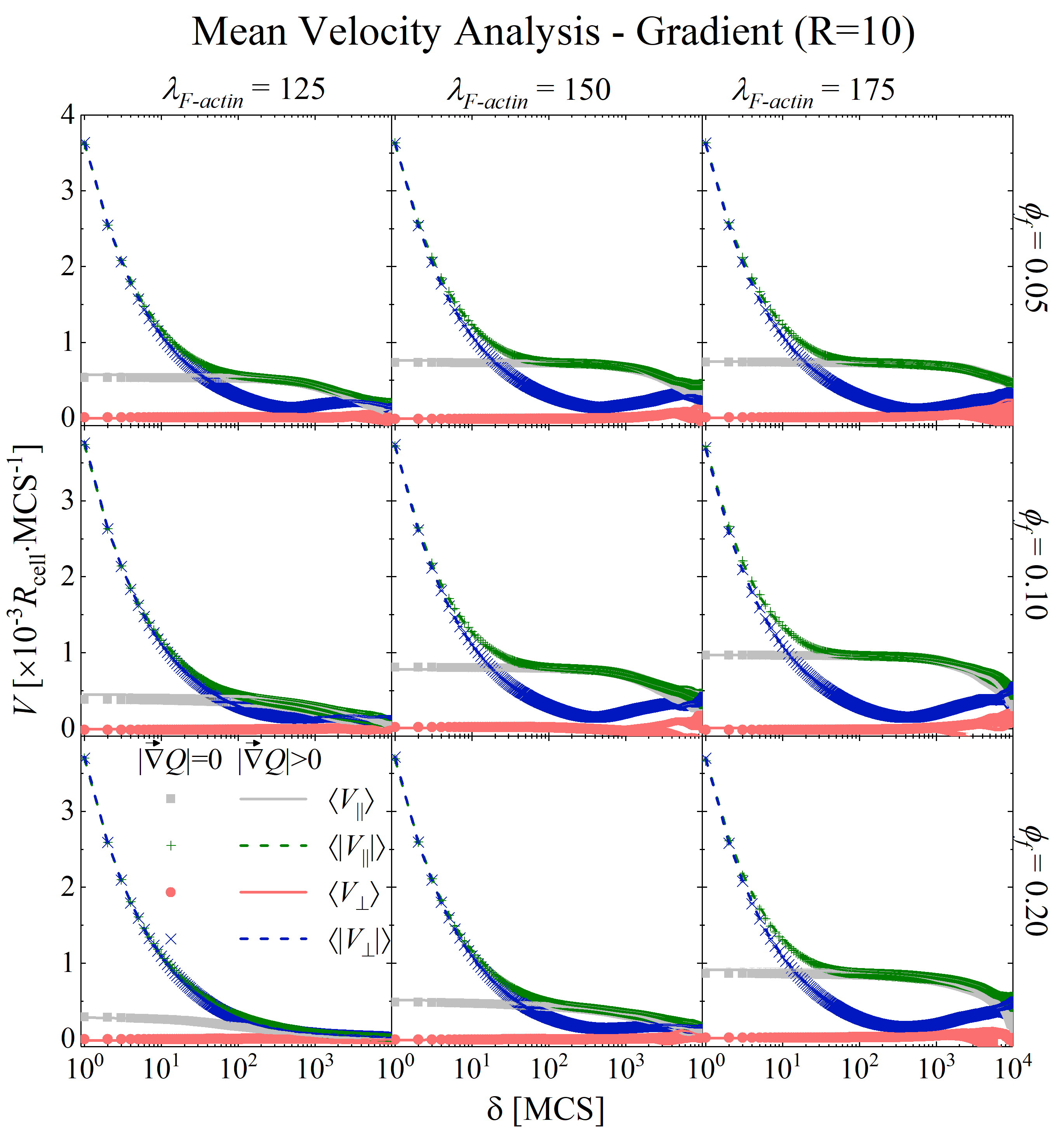}
    \caption{{\setstretch{1.0} Average mean velocity in parallel and perpendicular direction to polarization (represented with different colors) depending on Lamel protrusion coefficient $\lambda _{F-actin}$ (graph columns) and on the relative Lamel volume $\phi _f$ (graph rows), for two cases: \textbf{1)} environment is isotropic $\vec{\nabla} Q=0$ (points), and \textbf{2)} the cell is exposed to an external chemical gradient $\vec{\nabla} Q>0$ (lines). The parallel velocity $\langle V_\parallel \rangle$ (gray) shows little difference between the two cases for every parameter set, suggesting that the external gradient does not increase cell's drift speed. $\langle |V_\parallel |\rangle$ (green) and $\langle |V_\perp|\rangle$ (blue) also do not show significant differences between the two cases, which means the external gradient did not affect the diffusive dynamics as well.}}
    \label{fig:velfieldfig}
\end{figure}

We show in Fig. \ref{fig:velfieldfig} that the drift speed (defined in Eq. \ref{eq:VD}) does not change appreciably when the gradient is active $\vec{\nabla} Q>0$ compared to the isotropic case $\vec{\nabla} Q=0$. The same applies for the average of the absolute mean velocities $\langle |V_\parallel |\rangle$ and $\langle |V_\perp|\rangle$ for all $\phi_f$ and $\lambda_{F-actin}$ varied. Based on the observed data, we concluded that a slight alteration in the parallel velocity is unlikely to be the primary cause of the chemotactic response that is observed. The next step is to analyse polarization, as it can promote chemotaxis if a reorientation dynamics exists.

\subsubsection{Chemotaxis does not act as an external force.}

\begin{figure}{htbp}
    \centering
    \includegraphics[scale=0.7]{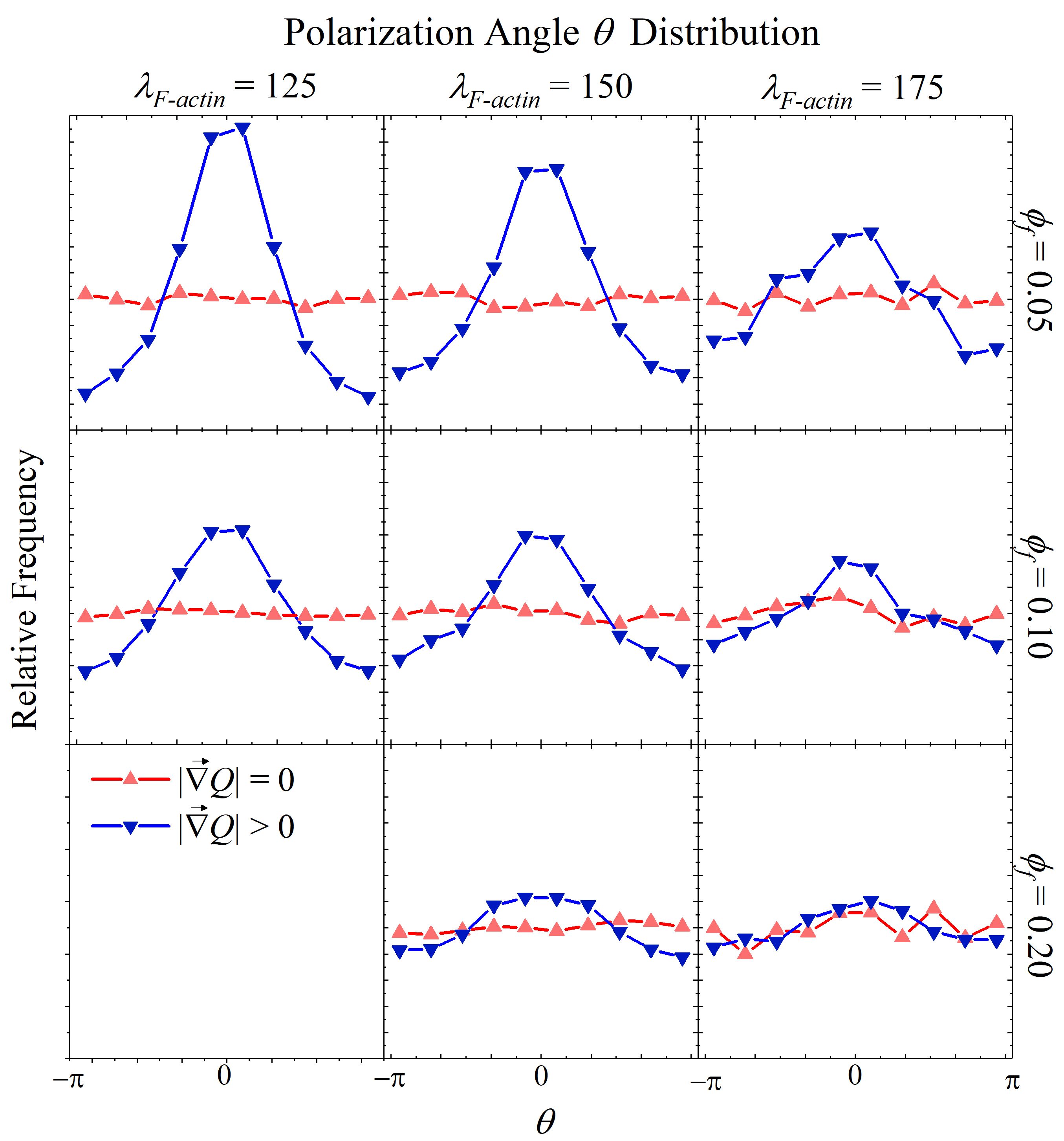}
    \caption{{\setstretch{1.0}Angle $\theta$ of the polarization vector in respect to the $x$-axis for different values of $\lambda_{F-actin}$ (graph rows) and $\phi_f$ (graph columns). Since the external field is linear in the $x$-coordinate, $\theta$ is also the angle between the polarization vector the chemical gradient. For $|\vec{\nabla} Q|>0$ (blue curves and points), $\theta$ distribution is narrow around $\theta=0$, while for $|\vec{\nabla} Q|=0$ (red curves and points)), $\theta$ distribution is flat. This figure indicates that the chemical gradient acts reorienting cell polarization.}}
    \label{fig:thetafig}
\end{figure}

To test the role of cell's orientation, we measured $\theta$ - the direction of cell's polarization (see Eq. \ref{eq:POL}) - throughout the simulations and made a histogram to check cell alignment with the field gradient. Fig. \ref{fig:thetafig} shows that the polarization angle $\theta$ relative to the $x$-direction has a flat distribution in the absence of external field, in contrast to a peaked distribution in the direction of the gradient when external field is present. The results in Figs. \ref{fig:velfieldfig} and \ref{fig:thetafig} suggest that the cell's migration dynamics in respect to polarization is preserved during chemotactic response, a topic we will further explore later. For now, we can conclude that our chemotaxis mechanism acts as guidance for cell polarity, not as an external force acting over cell's center of mass.

\subsubsection{Trajectories and MSD show long time ballistic regime in the presence of an external field}
\label{TRAJ}
\par

To numerically characterize our cells' kinetics, we employ MSD - Mean Square Displacement measurements and fit procedures based on mathematical models for cell dynamics. We also use MSD fit results to convert our data to natural units, allowing comparison between different computational cell models and experiments.

\begin{figure}{htbp}
    \centering
    \includegraphics[scale=0.7]{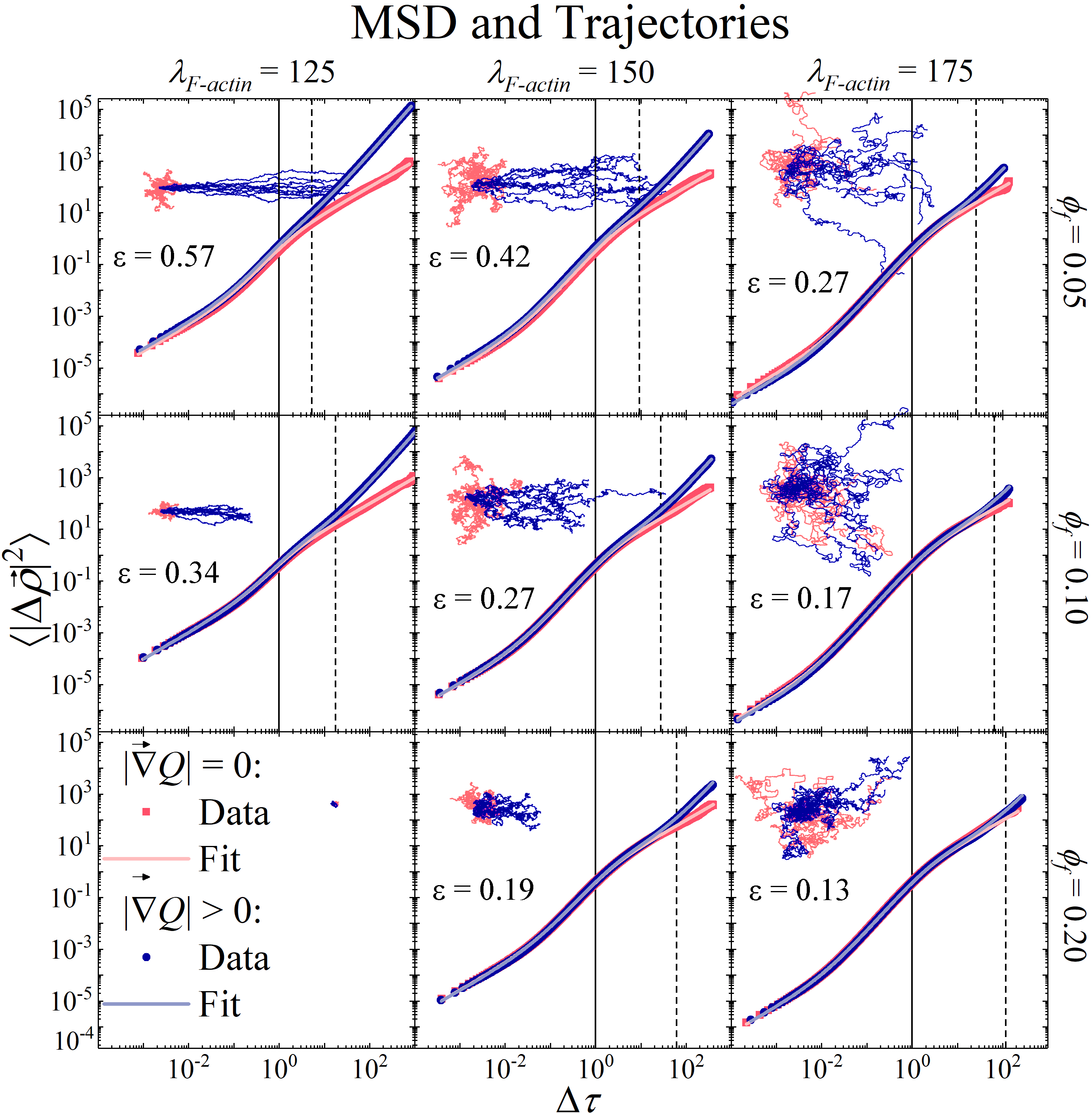}
    \caption{{\setstretch{1.0}Mean Square Displacement simulation data, fit curves and corresponding trajectories. Natural units for space and time are such that $\Delta \tau = \Delta t/P$ and $\langle \Delta\rho ^2\rangle = \langle \Delta r ^2\rangle (1-S)/2DP$. The solid vertical lines divide the persistent regime and the long term diffusive regime ($\Delta \tau =1$). The dashed vertical lines divide the long term diffusive regime and the long term ballistic regime (only valid when $|\vec{\nabla} Q|>0$). Dashed lines occur at $\Delta \tau = t_{taxis}/P$. Notice the trajectories for the set $\lambda_{F-actin}=125$ and $\phi_f=0.2$, they are the small dot beside the legend. We do not show the MSD for this specific set because it is impossible to find its natural units using Eq. \ref{eq:MSD} fit. We also show the chemotaxis efficiency $\varepsilon$ results for each case below the trajectories. In the Supplementary Materials \cite{Pedro2023supp}, Table S1 shows all results for red points (no chemotaxis), and Table S2 shows all results for blue points (with chemotaxis).}}
    \label{fig:msdfig}
\end{figure}

The MSD curves from experiments show three kinetic regimes for cell migration when external field gradient $|\vec{\nabla} Q|$ is zero: short time interval diffusion, intermediary time interval ballistic-like, and long time interval diffusive \cite{Thomas2019,Fortuna2020}. The present simulations show the same pattern, as shows Fig. \ref{fig:msdfig} (red dots). On the other hand, a fourth ballistic regime appears for time intervals long enough if an external field gradient $|\vec{\nabla} Q|>0$ is present. The MSD can be fit using the equation

\begin{equation}
   \langle  |\Delta\vec{ r}|^2 \rangle = 2D(\Delta t - P(1-e^{-\Delta t/P})) + \frac{2DS}{1-S}\Delta t + B \Delta t^2 \;\;\; ,
   \label{eq:MSDfit}
\end{equation}
where the first term is the Fürth equation - solution of the Langevin equation - with diffusion constant $D$ and persistent time $P$. The second term linear in $\Delta t$ accounts for the first diffusive regime, with diffusion constant $\frac{DS}{1-S}$, as proposed in Ref \cite{Thomas2019}. The third term accounts for the external gradient response, where $V_T = \sqrt{B}$ is the terminal speed, as defined in Eq.\ref{eq:VD}. In  Supplementary Materials \cite{Pedro2023supp}, we show a step-by-step fitting process of this MSD curve, and Tables S1 and S2 show all numerical results.

Taxis time $t_{taxis}$ is defined as the  time interval that separates the third regime (long term diffusive) and the fourth ballistic regime associated to chemotactic response can be calculated using the formula
\begin{equation}
    t_{taxis} = \frac{2D}{B(1-S)} \;\;\; .
\end{equation}

 We end up with 4 characteristic time intervals: 
\begin{enumerate}
    \item $\Delta t = SP$: the time interval that separates the first diffusive regime from the intermediate ballistic (the product between $S$ and $P$ parameters from the MSD fit)
    \item $\Delta t = \delta_{opt}$: the time interval where the intermediate ballistic regime has the highest slope, we call it "optimal delta"
    \item $\Delta t = P$: the persistence time, which separates the intermediate ballistic regime and the long term diffusive regime
    \item $\Delta t = t_{taxis}$: the taxis time, which separates the long term diffusive regime and the long term ballistic regime associated with the chemotaxis response.
\end{enumerate}

If the MSD presents these 4 regimes in the correct order, we expect that 

\begin{equation}
    SP< \delta _{opt} < P < t_{taxis} \;\;\; .
\end{equation}

\begin{figure}{htbp}
    \centering
    \includegraphics[scale=0.5]{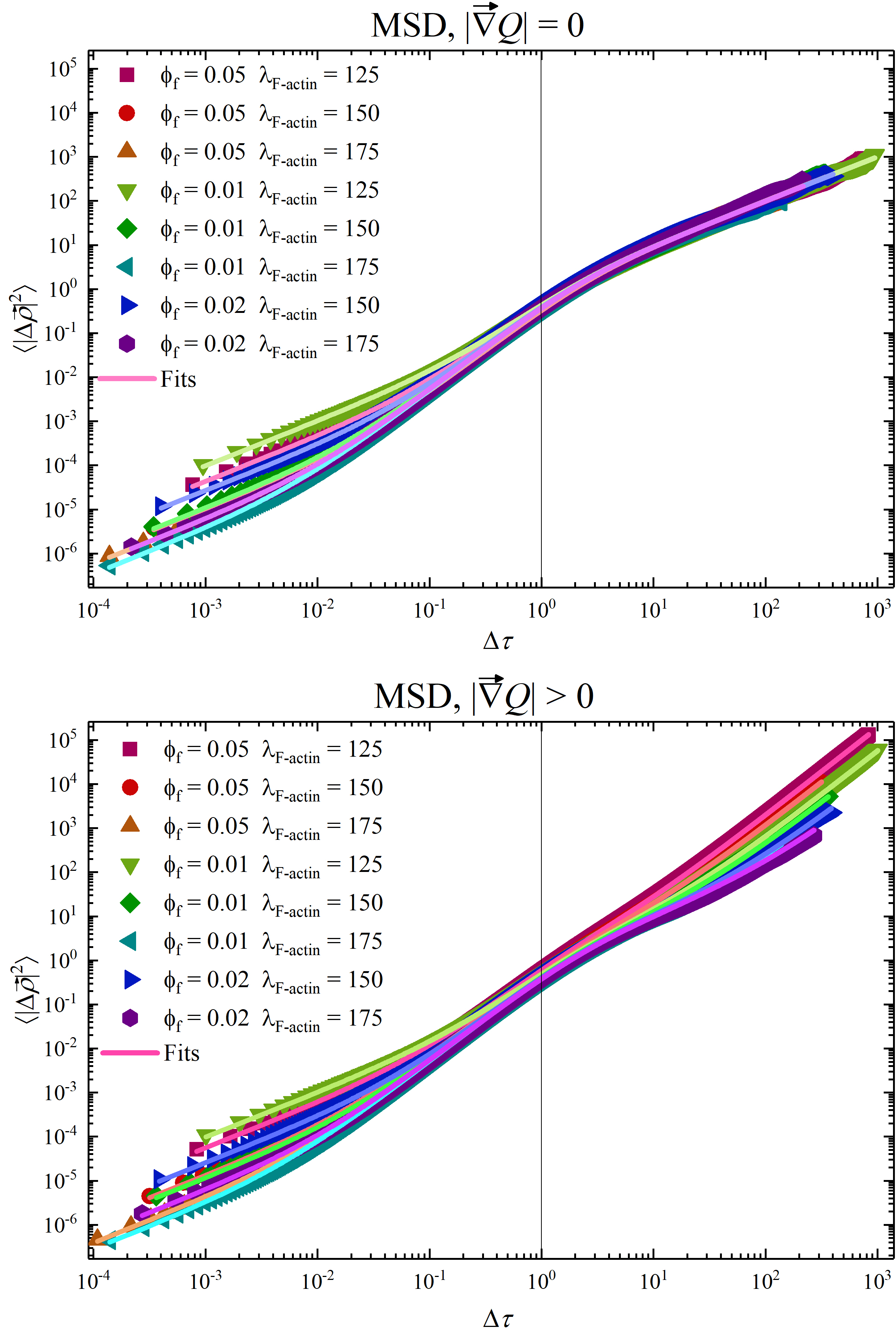}
    \caption{All MSD curves presented together in presence (upper panel) and absence (lower panel) of external field. Natural units lead to the collapse of the curves for $\Delta \tau =1$ (both cases) and for $\Delta \tau >1$ (only in the isotropic case).}
    \label{fig:MSDCollapse}
\end{figure}

The MSD curves in Figs. \ref{fig:msdfig} and \ref{eq:MSD} are presented in natural units of

\begin{equation}
     \langle  |\Delta\vec{ \rho}|^2 \rangle  =  \langle  |\Delta\vec{ r}|^2 \rangle \frac{1-S}{2DP}\;\;\;,\;\;\; \text{and}\;\;\;\;\;\; \Delta \tau = \Delta t/P \;\;\;,
\end{equation}
which collapse all curves into one if $S=0$ (Langevin limit) and $|\vec{\nabla} Q|=0$ (isotropy). Since each cell has a different S, curves separate for small time intervals, but remain united from $\Delta \tau=1$ forward. If cells have a terminal speed due to external field, then each cell will have a different value for $B$ as well, making curves separate after $\Delta \tau =1$. In Fig. \ref{fig:MSDCollapse} we show the curves collapsing for both cases: isotropic ($|\vec{\nabla} Q|=0$) and anisotropic ($|\vec{\nabla} Q|>0$). 

In the absence of external field, the long term diffusive regime is due to the change in polarization direction. In the presence of external field, the chemical gradient becomes a preferential direction for polarization to align. Since cell moves preferentially in the direction of polarization, a ballistic regime appears for very long time intervals i.e. $MSD \propto\Delta \tau^2$. Fig. \ref{fig:msdfig} also shows some typical trajectories for each set of parameters, in the presence (blue) and absence (red) of external chemical field to illustrate the change in behavior due to the action of the external field.

\subsubsection{Chemotactic efficiency increases for lower $\lambda_{F-actin}$ and lower $\phi_f$}
\par

\begin{figure}{htbp}
    \centering
    \includegraphics[scale=0.5]{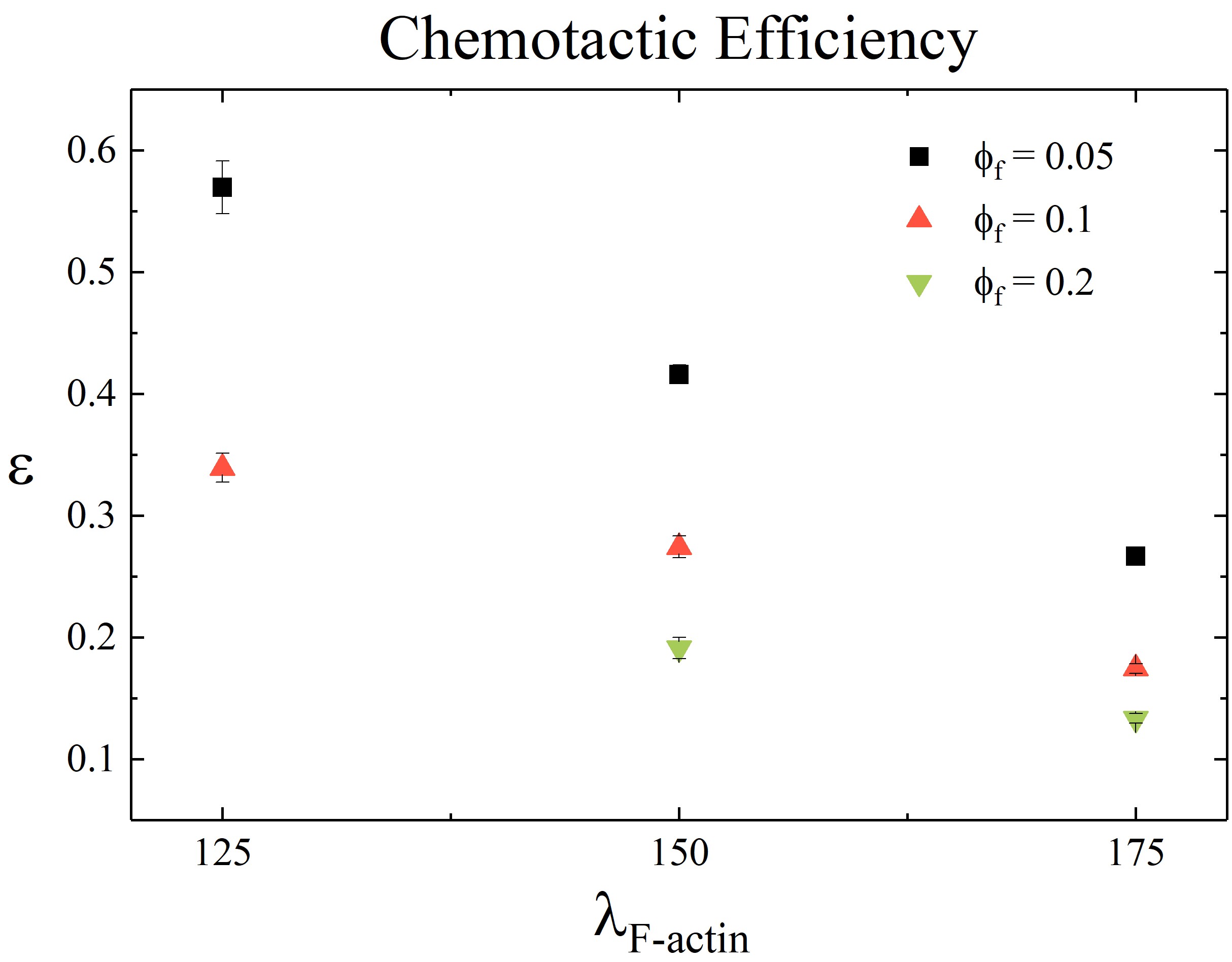}
    \caption{Relation between chemotaxis efficiency $\varepsilon$ and parameters $\lambda_{F-actin}$ and $\phi_f$. Simulation set with $\lambda_{F-actin}=125$ and $\phi_f=0.05$ does not present enough movement, so the metrics do not apply.}
    \label{fig:ChemEff}
\end{figure}

To determine how the simulation parameters impact the chemotactic response, we measure the chemotactic efficiency $\varepsilon$, defined in Eq. \ref{Eq:ChemEFF}. Chemotactic efficiency is the ratio between the terminal speed, acquired from MSD fits, and the drift speed, acquired via averaging the velocity component parallel to polarization. We show in Fig. \ref{Eq:ChemEFF} the relation between $\varepsilon$ and the studied parameters: protrusion coefficient $\lambda_{F-actin}$ and relative Lamel volume fraction $\phi_f$. 

The reason why $\varepsilon$ decreases with $\phi_f$ is related to Lamel behavior as a function of its size. As $\phi_f$ increases, Lamel spreads out more around Cyto, hindering polarization and reducing the relative effect of newly created Lamel voxels. 

$\lambda_{F-actin}$ regulates the protrusive force from F-actin gradient in the interface between Lamel and Medium as given by Eq. \ref{Eq:Eprotrusive}. Grater $\lambda_{F-actin}$ means more copies of Lamel voxels over the surrounding Medium in the $z=1$ plane. As a consequence, a newly created Lamel voxel back in Cyto will likely disappear (replaced by a Cyto voxel again) giving room to another Lamel copy towards Medium due to protrusion strength. In other words, higher $\lambda_{F-actin}$ makes Lamel a more stable structure wherever it is pointing to. Therefore, the Lamel creation mechanism has low effect over strong and stable Lamel. We conclude from this result that our cell presents a trade off between mobility in the absence of external gradients and chemotactic efficiency. This  result applies for our simulation model. It remains to be verified in experiments.

\subsubsection{mVACF shows a remaining correlation for long time intervals in the presence of external field, and displacement autocorrelation is negative for small time intervals}
\par

\begin{figure}{htbp}
    \centering
    \includegraphics[scale=0.7]{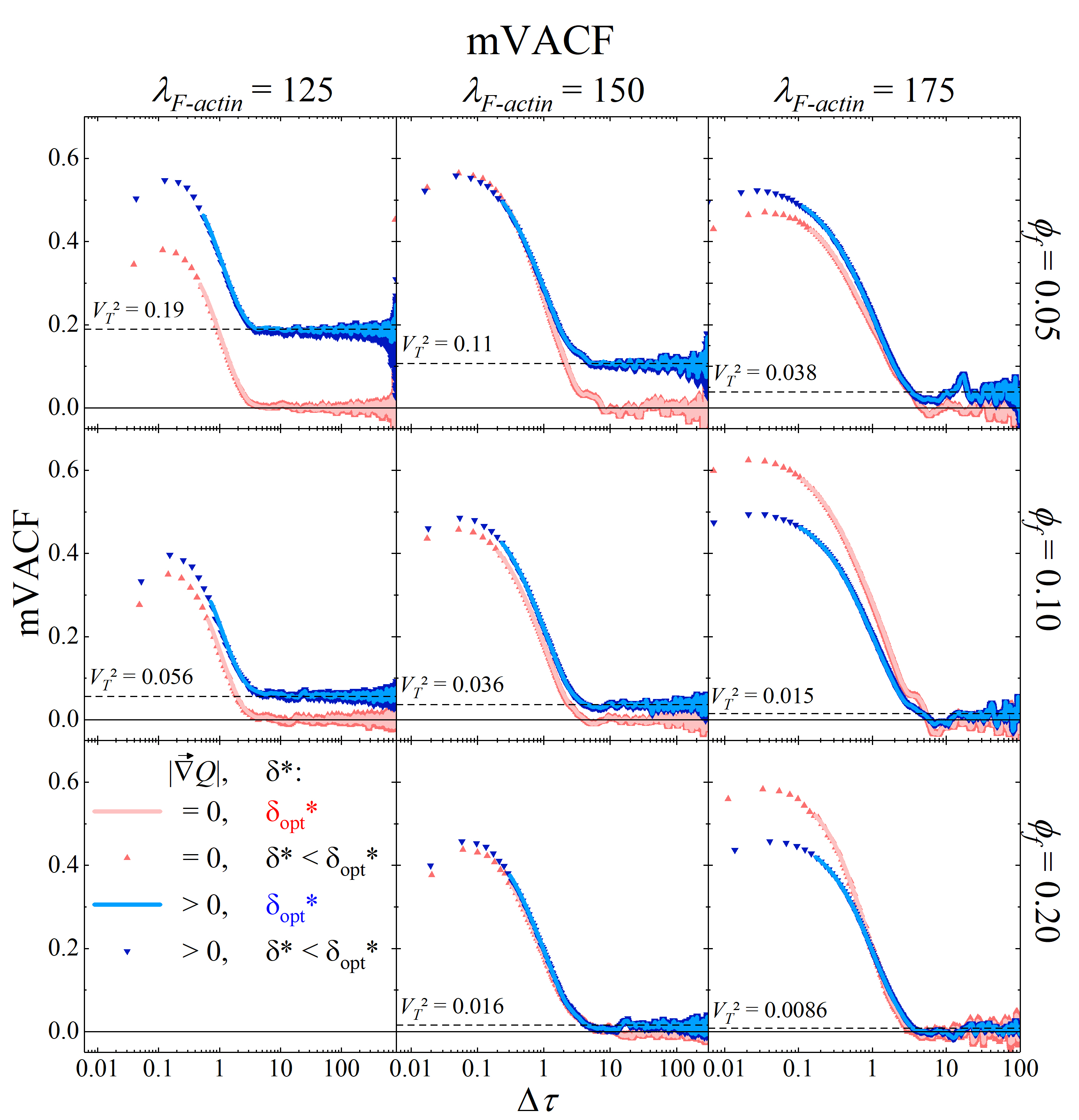}
    \caption{{\setstretch{1.0}Mean Velocity Autocorrelation Function. Symbols stand for mVACF using $\delta^*=50/P$. Straight lines represnt mVACF using $\delta _{opt}^*$ (in natural units) where $ \Delta \tau= \delta _{opt}^*$ corresponds to time interval that yields the maximum slope of the MSD in the persistent regime. We remark that  $\delta^*=50/P<\delta _{opt}^*$. For $\Delta \tau>0$, the $\delta^*=50/P$ curves (dark hue symbols) show that mVACF has a maximum. All straight lines (light hues) start after the peak showed by the symbols. As $\Delta \tau \rightarrow \infty$, mVACF goes to zero in the abscence of chemical gradient ($|\vec{\nabla} Q|=0$), or to a finite value in the presence of chemical gradient ($|\vec{\nabla} Q|>0$). In fact, the terminal speed squared $V_T^2$ obtained from the MSD fit coincide with the mVACF's rest value for large $\Delta \tau$ (in natural units).}}
    \label{fig:mvacffig}
\end{figure}

As we show in Fig. \ref{fig:mvacffig}, mVACF provides further details on the migration memory loss over time. Starting from $\Delta \tau= \delta^*$, as $\Delta t$ increases, mVACF shows first a rise in correlation, reaching a maximum value and thence an exponential decay towards an asymptotic value for large $\Delta \tau$. All cases where $|\vec{\nabla} Q|=0$ have a zero asymptotic value, which means there is no remaining memory. However, For $|\vec{\nabla} Q|>0$, the asymptotic value is greater than zero and depends on cell parameters. Positive velocity autocorrelation for large $\Delta \tau$ indicates directed migration. We show that the asymptotic value coincides with the squared terminal speed $V_T^2$. mVACF can be calculated using different $\delta$ (time interval used to measure mean velocity), and it gives similar results if $\delta$ is not too small.

\par

mVACF behavior for small $\Delta \tau$ implies a negative autocorrelation and, hence, an additional dynamical term for cell velocity. This has been observed before in \cite{Thomas2019,Fortuna2020,Li_2008,Dieterich08,Li_2011}. Thomas 2019 proposed an explanation of a loss in measurement precision due to the higher weight of diffusive terms as compared to drift terms when $\delta \rightarrow 0$. However, this effect would also rule out the possibility of measuring drift speed for small values of $\delta$, as shown in Figs. \ref{fig:velfig} and \ref{fig:velfieldfig}.

\begin{figure}
    \centering
    \includegraphics[scale=0.5]{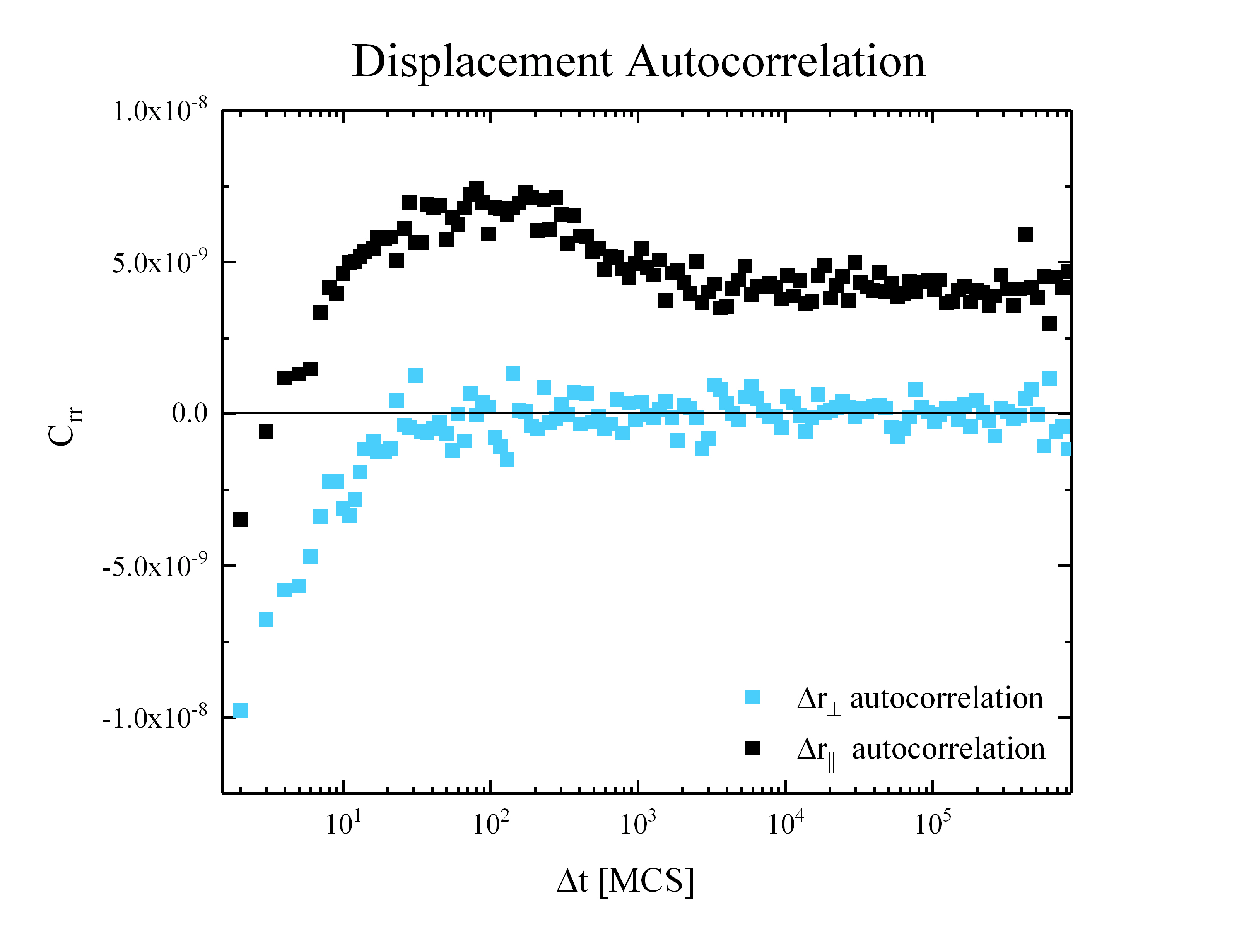}
    \caption{Displacement Autocorrelation Function versus the time interval $\Delta t$ for parameters $\phi_F=0.1$, $\lambda_{F-actin}=125$, without external gradient. The displacement parallel to polarization ($\Delta r_{\parallel}$) shows a positive correlation for large $\Delta t$, which confirms the presence of a drift in this direction. The displacement perpendicular to polarization ($\Delta r_{\perp}$) has zero correlation for large $\Delta t$, showing zero drift in this direction. Both displacements go to negative correlation as $\Delta t \rightarrow 0$, but $\Delta r_{\perp}$ falls faster.}
    \label{fig:Crr}
\end{figure}

Our hypothesis to explain this effect is that the cell dynamics for short time intervals yields a negative displacement autocorrelation function. To test this hypothesis, we measured $C_{rr}(\Delta t)$ for both parallel and perpendicular directions to polarization. Fig. \ref{fig:Crr} shows what we expected; the correlation is negative for low values of $\Delta t$. We lack a dynamical explanation for this effect and leave for future work.

\subsubsection{Cell displacement probability distribution are different in the laboratory and cell reference frames}
\par
\label{Volcanos}

In Section \ref{velocity1} we inferred that our cell's dynamics is preserved during chemotactic response. To test this hypothesis, we calculate cell displacement components in two reference frames: \textbf{1)} cell reference frame defined by the polarization and \textbf{2)} laboratory reference frame, defined by the lattice coordinates.

\begin{figure}{htbp}
    \centering
    \includegraphics[scale=0.4]{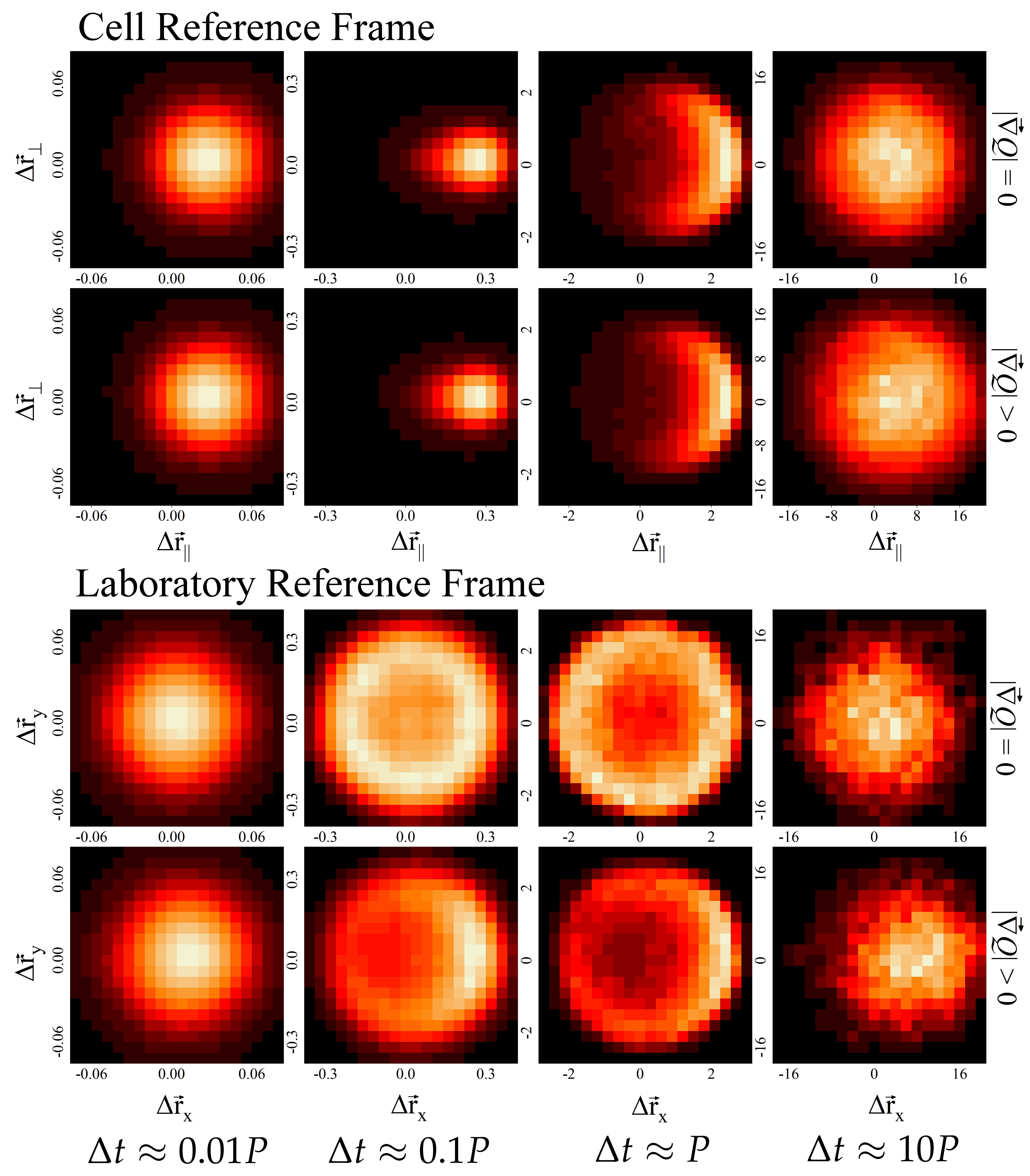}
    \caption{{\setstretch{1.0} The displacement distribution is the 2D histogram of the cell displacement in a given $\Delta t$. There are 4 columns of graphs, one for each $\Delta t \approx {0.01P,\; 0.1P,\; P,\; 10P}$, where $P$ is the persistence time. In the first 2 graph rows, the displacement is calculated in the cell reference frame defined by the cell polarization vector in the beginning of the step. The displacement relative to the polarization has coordinates $\Delta \vec{r}_{\parallel}$ (displacement parallel to polarization vector) and $\Delta \vec{r}_{\perp}$ (displacement perpendicular to polarization). In the 3rd and 4th graph rows, the displacement is calculated in the laboratory reference frame (Petry dish). The displacement relative to the Petry dish has coordinates $\Delta \vec{r}_{x}$ and $\Delta \vec{r}_{y}$. The cell parameters here are $\lambda _{F-actin}=150$ and $\phi _{f}=0.1$.}}
    \label{fig:vulcoes}
\end{figure}

We picked the simulation with parameters $\phi_f=0.1$ and $\lambda_{F-actin}=150$ to show the displacement distributions in Fig. \ref{fig:vulcoes}, considering four different time intervals $\Delta t$: $0.01P$ (fast diffusive), $0.1P$ (ballistic-like), $P$ (ballistic-like), and $10P$ (slow diffusive or ballistic depending on $\vec \nabla Q$ conditions) MCS. We also considered two reference frames: laboratory and cell reference frames, with and without the external field. The external field gradient orientation is aligned to the $x$-axis of the laboratory reference frame. In the cell reference frame, $x$-axis is the cell's polarization direction in the beginning of the step. 

In the cell reference frame, displacement distributions change depending on time scale $\Delta t$. For small $\Delta t$, distribution is Gaussian-like and centered at a positive value of the $x$-axis, indicating a movement with diffusion and a drift aligned to polarization. For $\Delta t$ close to persistence time, asymmetry increases, showing the correlation between polarity and displacement in these time scales. For time scales at the order of $10P$, displacement distribution becomes Gaussian-like again. When comparing cases with $|\vec{\nabla} Q|=0$ and $|\vec{\nabla} Q|>0$, we do not see significant differences, indicating that migration dynamics stays preserved when chemotactic response takes place.
 
In the laboratory frame of reference, displacement distributions also depend on time scale $\Delta t$. For short $\Delta t$, distributions are Gaussian-like, indicating again the diffusive noise. For $\Delta t$ close to persistence time, we found ring-like distributions, indicating that longer steps are more likely to occur. For time scales about $10P$, distributions become Gaussian-like again. Unlike the cell reference frame, the laboratory frame shows clear differences between the $|\vec{\nabla} Q|=0$ and $|\vec{\nabla} Q|>0$ cases. $|\vec{\nabla} Q|=0$ distributions are symmetric, while $|\vec{\nabla} Q|>0$ distributions are asymmetric due to the directionality of polarization angle $\theta$.







------------------------------------------------------

\section{Conclusion and discussion}

\subsection{Overview}

We presented a model that reproduces chemotaxis behavior in a linear, constant chemical field where the cell responds to the gradient by reorienting polarization rather than being pushed by it, with directional sensing, polarity reorientation and migration as individual, simultaneous processes. Directional sensing is achieved by measuring the chemical field concentration in the contact between the cell and the substrate, reorienting is achieved by a localized creation of Lamel, and finally, the F-actin's protrusion energy acting over Lamel pushes the cell in the direction of polarization.

With this model, we showed it is possible to separate the cell dynamics in a new coordinate system defined by the cell's polarization. This method allowed us to identify a diffusive behavior in cell velocity that is similar in all directions i.e. isotropic. Together with this diffusive noise, we found that the cell has a net velocity only in the polarization direction, which we called drift speed.

When submitting our simulated cell to an external chemical gradient, we find no significant change in drift speed. Then we calculated the distribution of the polarization direction, which is very different between the cases with gradient and without: the distributions show a maximum of polarization direction in the gradient direction. Hence, our model's main mechanism to chemotactic response is polarization reorientation.

Once understood the basic cell behavior, we proceeded with more characterization metrics. The MSD shows a long term ballistic regime when the cell responds chemotactically, and all curves show the same diffusive behavior for short time scales (not following Langevin models). Nonetheless, we fit all MSD curves with a modified version of the Fürth equation and found 4 distinct time intervals of interest: 

\begin{enumerate}
    \item $SP$: the time interval that separates the first diffusive regime from the intermediate ballistic regime,
    \item $\delta_{opt}$: the time interval where the intermediate ballistic regime has the highest slope,
    \item $P$: the persistence time, and
    \item $t_{taxis}$: the taxis time, which separates the long term diffusive regime and the long term ballistic regime associated with the chemotaxis response,
\end{enumerate}
where $S$ and $P$ are results from the MSD fits. We also extracted $V_T$ (terminal speed) from the MSD fits, which we then used to calculate chemotaxis efficiency and taxis time $t_{taxis}$. 

We defined a robust metric of chemotactic efficiency that can be applied to any cell that has a net speed in the direction of polarization (drift speed) and responds to the chemical gradient with chemotaxis, whatever the response mechanism may be. We showed that \textbf{1)} a larger and more spread out Lamel around Cyto lower the chemotactic efficiency due to reduced relative effect of the Lamel creation mechanism; and \textbf{2)} a strong and stable Lamel is less affected by the Lamel creation mechanism, also reducing chemotactic efficiency. Qualitatively, these results translate into biology as a trade off between cell polarization stability and its ability to respond to external chemical gradients. 

The mVACF showed a remaining correlation for long time intervals in the presence of the external chemical gradient, which is equal to $V_T^2$ (the terminal speed squared), and also showed that the correlation falls for small time intervals, indicating a negative autocorrelation in the dynamics of cell displacement, which we then verified.

At last, we measured the displacement distribution for different time intervals and used two reference frames of interest: the cell reference frame, defined by its polarization, and the laboratory reference frame, defined by the lattice, yielding different outcomes. For instance, when comparing chemotactic cells to non chemotactic cells, we only found a difference in the laboratory reference frame, due to directed migration. However, similar distributions were found in the cell reference frame, suggesting that the cell's dynamics relative to the polarization remains unchanged.

\subsection{Takeaways for Experimental Biologists}

Our model structure and the way we collected data were inspired in how a real experiment is conducted. The similarities enable us to make a few considerations on experiment design and data analysis regarding cell migration:

\begin{enumerate}
    \item Make sure the sampling time interval and experiment length cover all expected regimes of cell movement. To determine what is a small time interval and what is a large time interval, we suggest running a few trials before a full experiment set, to check if the Mean Square Displacement - MSD encompasses all expected regimes. As a starting point, the time it takes for a cell to make 3 to 5 successive full polarization reorientations indicates a large time interval; the time it takes for a single lamellipodium protrusion to happen is a good indicator of a small time interval. 
    \item Investigate the presence of a diffusive regime for very short time intervals. Good candidates are cells with frequent membrane fluctuation or frequent lamellipodia protrusions and retractions. Collective cell environments can also be a source of diffusive noise at short time intervals due to frequent interactions between cells. To check if the diffusive regime exists, plot the MSD in log-log scales and measure the MSD slope in short time intervals. A slope below 2 is a strong indicator of diffusive noise.
    \item If cell movement presents a diffusive regime for short time intervals, instantaneous velocity is not well defined. To circumvent this, we suggest using the optimal velocity, which is the mean velocity measured for the time interval at which the MSD has its highest slope. To characterize cell motility, we suggest using drift speed, as this metric is well defined with or without diffusive noise, provided that both velocity and polarization are measured.
    \item The starting point of cell migration characterization is plotting and fitting the MSD curve. Besides the detection of different regimes, the MSD fit provides parameters such as diffusion, persistence, and terminal speed (with chemotactic response). These parameters can be used to compare cell motility results from different experiments, and also with computational model results. We provide a step-by-step procedure to fit a case of 4 regimes, which requires 4 parameters. The quality of the fit depends directly on whether all regimes of movement were covered (refer to item 1), and on the number of replicas (we used 10 replicas).
    \item The cell reference frame, defined by cell polarization, is not a new concept. But using it to project metrics like cell displacement is still not heard of. For example, we demonstrated how to use cell displacement distribution in cell reference frame and laboratory reference frame to unveil the mechanism behind cell movement and chemotactic response, see Fig. \ref{fig:vulcoes}.
    \item Cell polarization unlocks the use of sophisticated metrics and circumvent the problems with instantaneous velocity. A proxy to polarization is any vector correlated to the optimal velocity: cell shape, average gradient of some molecule like actin, average force exerted from the substrate, or the displacement of the nucleus from the rest of the cell body as we did in our simulations. See \cite{Thomas2022} for a detailed guide on studying polarization correlation to cell displacement.
\end{enumerate}

\subsection{Final Thoughts}

We believe that our model can be applied to other taxis processes, since they are similar in a simplistic mathematical standpoint: the cell senses some anisotropy in the environment and adapts its migration direction. The simulation allows changes in the sensing mechanism, making it possible to study receptors' density, saturation, activation, delayed response, and forced protrusion. We are left with the task of building a pure mathematical model that can describe this specific dynamics, from which we should derive Eq. \ref{eq:MSD} with the extra ballistic term. The next modification to be made over this simulation is to accomplish collective migration to study wound healing or metastasis, or keep it single cell and apply to a immunologic response problem.

To be useful for biological cells, we must verify whether this three-step process of chemotaxis yields constant drift speed in experiments. This can be achieved by measuring polarization and plotting displacement probability density functions in the cell and lab reference frames.

Our findings have a particular relevance to collective migration, as varying degrees of collective polarization and lamellipodium production can result in distinct  behaviors. We thence expect that our paper will be useful for developing collective cell migration simulations in biological phenomena as in wound healing, metastasis and immune response.

\section*{Acknowledgments}
This work has received support from Brazilian agencies CAPES - Coordenação de Aperfeiçoamento de Pessoal de Nível Superior, and CNPq - Conselho Nacional de Desenvolvimento Científico e Tecnológico. We thank UFRGS - Universidade Federal do Rio Grande do Sul, and PPGFIS - Programa de Pós Graduação em Física, for providing the infrastructure and administrative support that enabled this project.

\bibliographystyle{unsrt}
\bibliography{CellMig}

\end{document}